\documentclass[
    ,final            
  ]
  {aipproc}

\layoutstyle{8x11double}

\begin{document}

\title{Structured jets and VHE emission of blazars and radiogalaxies}


\author{Fabrizio Tavecchio}{
} \author{Gabriele Ghisellini}{ address={INAF--Osserv. Astron. di Brera,
via E. Bianchi 46, 23807 Merate, LC, Italy} }

\begin{abstract}
Recent observations in the TeV band challenge the simplest models
developed to describe the overall emission of blazars and
radiogalaxies. In particular, the observation of variable TeV emission
from M87 and the fast variability shown by PKS 2155-304 challenge the
standard framework. We discuss how the existence of a radial structure
in the sub-pc scale jet, with faster a component (``spine'' or
``needles'') embedded in a slower layer can explain the basic
phenomenology of these sources.
\end{abstract}

\maketitle


\section{Introduction: the structured jet model}

Several observational and theoretical clues suggest that jets in
extragalactic sources can be structured, with a fast core ({\it
spine}) surrounded by a slower {\it layer}. Among the evidences coming
from observations we recall the direct radio imaging of
structures in the innermost regions of the close-by BL Lac objects
and radiogalaxies [1,2,3]
and the two-velocity structure required to unify FRI radiogalaxies
and BL Lacs [4]. Theoretically, a structure
in the jet alleviates some problems related to the models of
TeV BL Lacs [5]. Recent simulations of jet
formation support the existence of a spine--layer structure already in
the initial phases of the jet propagation [6,7].

The existence of a velocity structure has a strong impact on the
observed emission properties of jets. Specifically, the radiatively
interplay between the layer and the spine amplifies the inverse
Compton emission of both components [8,9]. Indeed, both components
will see the emission of the other amplified because of the relative
speed. This ``external'' radiation contributes to the total energy
density, enhancing the emitted inverse Compton radiation. Depending on
the parameters, this ``external Compton'' (EC) emission can dominate
over the internal synchrotron self-Compton (SSC) component that,
especially in TeV blazars, is depressed because scatterings mainly
occur in the Klein-Nishina regime.

An important point to consider is that the emission from the layer is
beamed within the angle $\theta _l\sim 1/\Gamma _l$ (where $\Gamma _l$
is the bulk Lorentz factor of the layer), larger than the
corresponding angle for the spine, since $\Gamma _s>\Gamma _l$. This
implies that the layer can be seen at relatively large viewing angles
for which, instead, the emission from the spine is severely
depressed. A direct prediction of this fact is that, besides blazars
(dominated by the spine), also misaligned jets in radiogalaxies could
be relatively strong $\gamma$-ray emitters, dominated by the layer [5].

In the following we use the spine-layer scenario to interpret the
observed TeV emission from M87 and the challenging rapid variability
(down to few minutes) recently observed from some TeV blazars.

\section{TeV emission of M87}


\begin{figure}
  \includegraphics[height=.42\textheight]{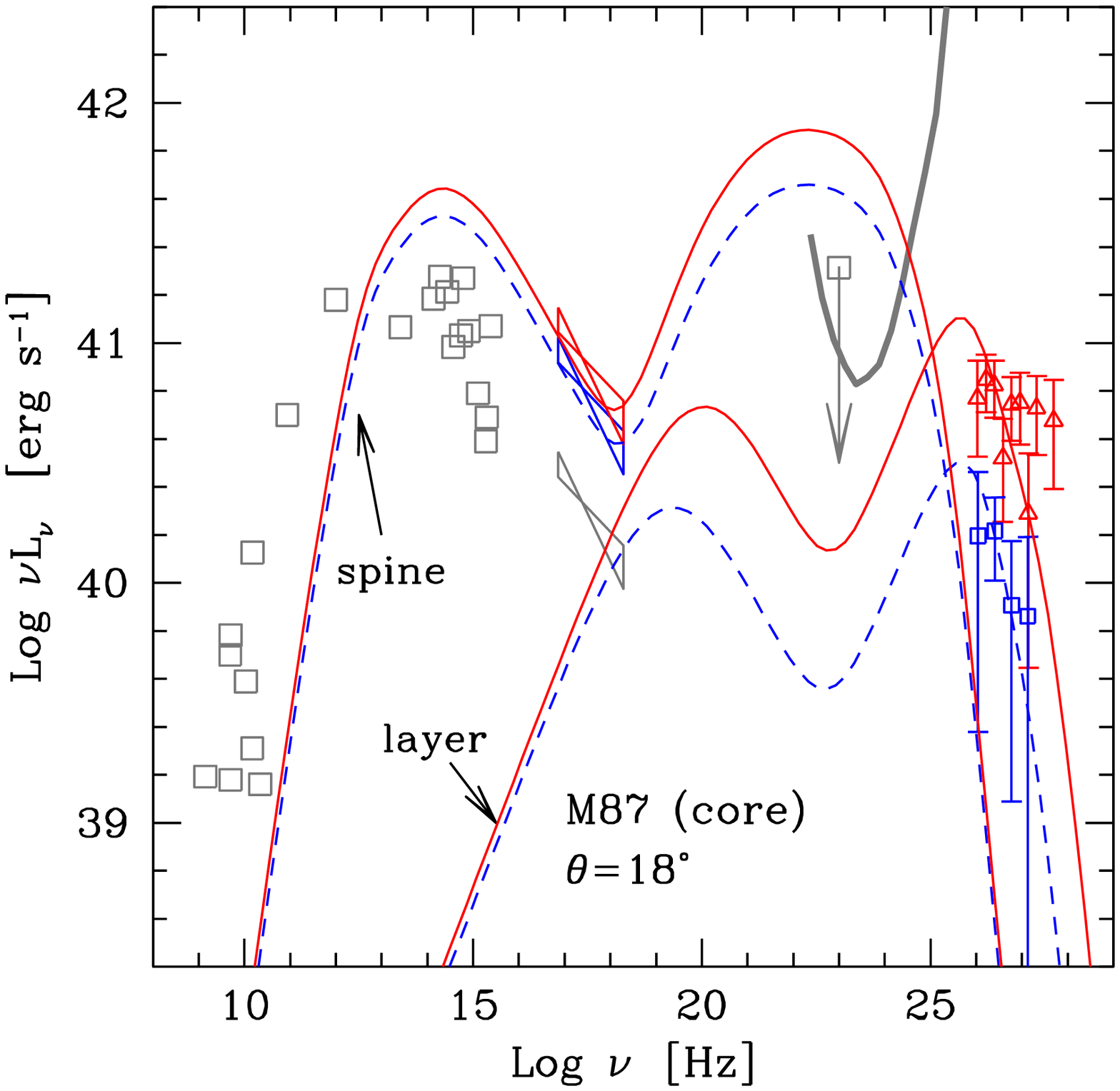}
\includegraphics[height=.42\textheight]{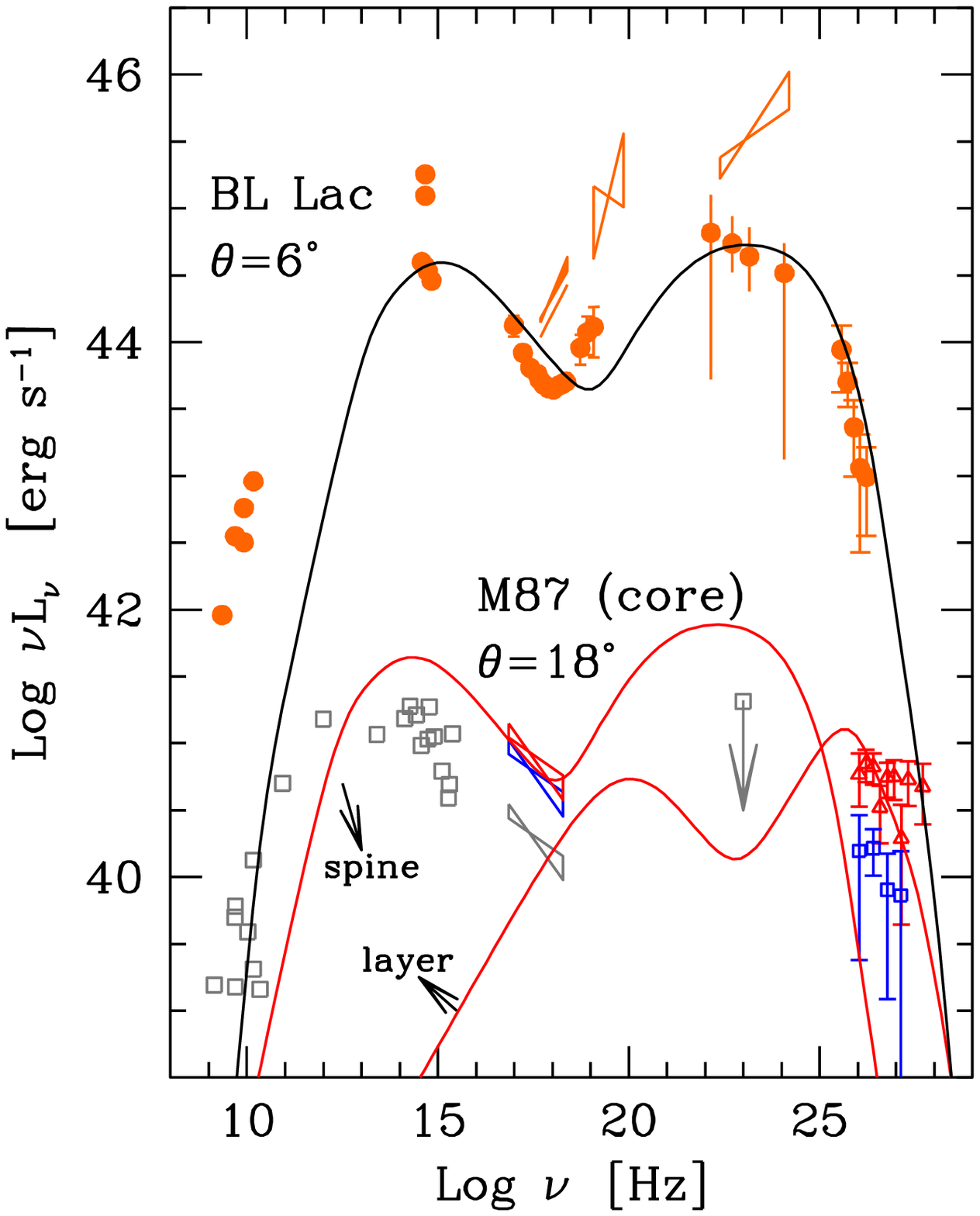}
\vspace*{-0.9 truecm}
  \caption{{\it Left:} SED of the core of M87 (open squares) together
with the H.E.S.S. spectra taken in 2004 (open squares) and 2005 (open
triangles), from [10]. The lower bow--tie reports the X--ray spectrum
as measured by {\it Chandra} in 2000. We reproduced the increased
X--ray emission of 2004 and 2005 (higher bow-ties) assuming the same
slope and a larger normalisation. The lines report the emission from
the spine and from the layer for the two states (dashed: 2004; solid:
2005). For comparison, the thick line indicates the sensitivity of
{\it Fermi} ($5\sigma$, 1 year, converted in luminosity assuming the
distance of M87). {\it Right:} The upper line reports the spine
emission computed for a viewing angle of 6 deg, compared with data
obtained for BL Lac (filled symbols). The corresponding emission from
the layer is well below that of the spine and is not reported for
simplicity. From [20].}
\label{fig1}
\end{figure}

The nearby (16 Mpc) radiogalaxy M87 has been discovered as a TeV source
by the HEGRA array [9]. Subsequent observations by
H.E.S.S., VERITAS and MAGIC confirmed the emission and showed that the
TeV flux is variable, both on short ($\sim$2 days) and long (years)
timescales [10,11,12]. Though the limited spatial resolution of Cherenkov
telescopes prevents to localize the emission region, the short
variability timescale allows us to rule-out models predicting TeV
emission from the kpc-scale jet [13].

Among the possible scenarios advanced to explain the observed
emission, that considering the emission from the peculiar knot HST-1,
located at 60 pc (projected) from the core [14,15] was supported by
the apparent correlation of the measured TeV flux and the X-ray
emission of HST-1 as measured by the monitoring of {\it
Chandra}. X-ray measures are difficult, since the separation of the
core and HST-1 is at the limit of the capabilities of {\it
Chandra}. Recent observations, showing an increase of the TeV flux not
accompanied by a corresponding increase of the X-ray brightness of
HST-1 [11] though not completely ruling out the
connection between TeV emission and HST-1, open the possibility that
the VHE emission originates in the core.  Moreover, the short
variability timescales seem difficult to reconcile with the size of
HST-1 without invoking some special geometry at the shock [15,16].

Models in which the emission region is located close to the core do
not have problems in explaining the short timescale
variability. Neronov \& Aharonian [17] proposed that the TeV emission
comes from relativistic particles accelerated by magnetic fields close
the central supermassive black hole. A more direct possibility is that
the emission comes from the slightly misaligned ($\theta \simeq 20$
deg) inner jet [18]. However, it can be shown that a simple
homogeneous synchrotron-SSC model fails in reproducing the entire
spectral energy distribution of the core of M87, mainly because of the
large separation of the synchrotron and SSC peaks in the SED [19,20],
requiring unreasonably large Doppler factors ($\delta\sim 500$).

The spine-layer scenario can easily overcome this problem [20].  In
our model (Fig.\ref{fig1}) the emission from the spine (with bulk
Lorentz factor $\Gamma _s=12$) accounts for the low-energy emission,
from radio to the GeV band, while the VHE component is produced by the
layer (with bulk Lorentz factor $\Gamma _l=4$). To produce TeV photons
we have to assume that electrons in the layer are highly relativistic
and thus the corresponding synchrotron radiation peaks at relatively
high frequency, above the X-ray band, where the SSC radiation of the
spine dominates. The SSC emission in the layer occurs mainly in the KN
regime and thus is strongly suppressed.  The high-energy peak of the
layer is thus largely dominated by the EC component. The model clearly
predicts that during states of high X-ray flux M87 should be a bright
source at GeV energies, a prediction that {\it Fermi} should easily
test.

\begin{figure}
  \includegraphics[height=.4\textheight]{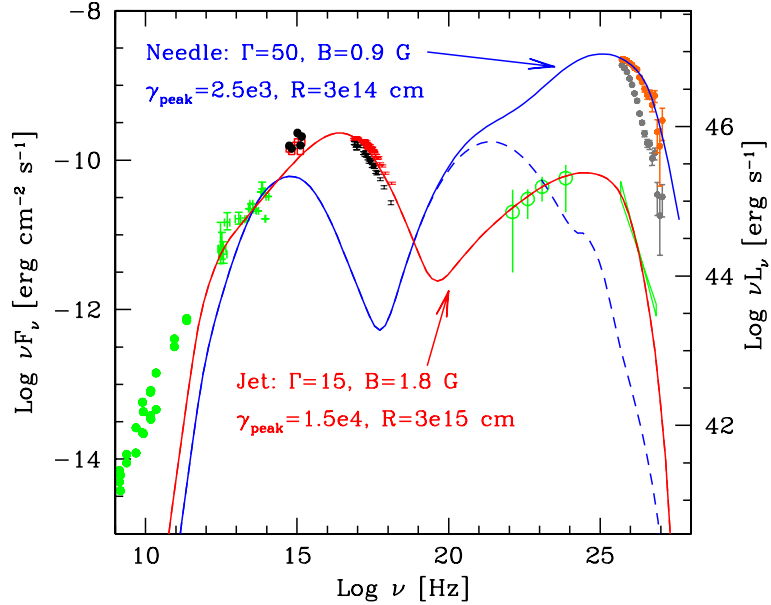}
  \caption{The SED of PKS 2155-304. Observed TeV data from HESS
  correspond to the flare of 2006 July 28 [22].  Red points report the
  TeV spectrum corrected for the extragalactic absorption (see text
  for details). X-ray and optical data are not strictly simultaneous
  to the TeV ones, but corresponds to 2 and 4 d later. Other symbols
  are archival data. The dashed line corresponds to the flux produced
  by the needle if we neglect the radiation energy density produced by
  the rest of the jet. The radiation energy density seen by the jet
  due to the needle emission is assumed to be negligible. From [27].}
\label{fig2}
\end{figure}

An important constrain that such a model has to satisfy is that the
SED seen by an observer located at small angle should display a shape
similar to that of known blazars. For comparison, in Fig.\ref{fig1}
the filled symbols report the observational data for the prototype of
BL Lac objects, BL Lac itself. The upper solid line is the SED of M87
measured by an observer locate at $6$ deg with respect to the jet
axis. At small angles the jet is dominated by the beamed emission of
the spine, while the less amplified layer component provides a
negligible contribution. Though we do not intend to exactly reproduce
the SED of this particular blazar, one can see that the model follows
quite well the observed data.

A direct prediction of the structured jet model is that {\it Fermi}
(and possibly Cherenkov telescopes) should detect other radiogalaxies,
probably those with the (inner) jet only slightly misaligned with
respect to the line of sight. Other possibilities to produce
detectable fluxes of $\gamma$-rays from radiogalaxies, not critically
dependent on the viewing angle, include the emission of the kpc scale
jet [13], of the hotspots [21] or of the lobes. The most direct way to
distinguish the origin of the high-energy emission is through the
variability. Fast ($\sim $days) variability would directly exclude
possibilities involving large scale regions, indicating that the
emission originates in the most compact regions of the radiogalaxy
(jet or BH).

\section{Rapid TeV variability of PKS 2155-304}

In summer 2006 the TeV BL Lac PKS 2155-304 showed a period of extreme
variability in the TeV band [22]. During the night of July 28 well
resolved flares varying on timescales of 200 seconds were observed
(similar variations have been also observed in Mkn 501, [23]). In this
phase the source was very active at VHE, reaching observed
luminosities of $10^{47}$ erg/s (to be compared with more typical
luminosities of $\sim 10^{45}$ erg/s). Such short variability
timescales are difficult to explain in the standard framework
[24]. Indeed, in the widely assumed internal shock scenario variations
should last for times larger than the timescale associated to the
black hole, $t_{\rm var}>R_s/c\sim 1.4 M_9$ h, where $R_s$ is the
Schwarzschild radius of the BH. On the other hand, if the source is a
moving sphere, Doppler factor as large as $\delta =50-100$ are
required in order to keep the compactness of the source to acceptable
values [25,26].

Also in this case a structured jet offers a good way to reproduce
these states [27] without a radical change of the theoretical
framework (Fig.\ref{fig2}). In this version the role of the spine is
played by a ``needle'', a very compact (size $\sim 3\times 10^{14}$
cm) region inside the ten times bigger ``normal'' jet, responsible for
the emission observed during most of the time. Though the needle is
characterized by a high bulk Lorentz factor, $\Gamma =50$, the total
power needed to reproduce the observed emission does not exceed that
carried by the normal jet.

As in the model for M87, a crucial role is played by the radiative
interplay between the needle and the normal jet. As can be seen in
Fig.\ref{fig2}, the bulk of the IC emission from the needle is
produced through the scattering of the photons emitted by the jet,
whose energy density is amplified by the high relative speed. Note
that the contribution of the needle at other frequencies is minor,
consistently with the small variability observed during the active
phases, especially at X-ray frequencies (e.g. Costamante, these
proceedings). Such a scenario could thus easily explain the so-called
``orphan flares'', TeV flares without a counterpart in the X-ray band
[28]. This also offers an effective way to test our scenario through
multifrequency observations, especially in the crucial X-ray and UV
bands. If ultrafast variations on the TeV band are not accompanied by
corresponding variations at lower frequencies, then the ``needle-jet''
model (or, more generally, any model considering more than an emission
region) is preferred.

A consequence of the ``needle-jet'' model is that the power of the jet
is dominated by the electrons (and protons) carried by the ``normal''
flow, while the magnetic field provides a negligible role (contrary
the conclusions of [24]).


\begin{theacknowledgments}
We tkank Laura Maraschi for useful comments
\end{theacknowledgments}


\end{document}